\documentclass{PoS}
\usepackage{float}
\usepackage{lineno}

\title{Determination of the extragalactic background light spectral energy distribution with H.E.S.S.}

\ShortTitle{Determination of the EBL SED with H.E.S.S.}

\author{\speaker{Matthias Lorentz}\\
        DSM/Irfu, CEA Saclay, F-91191 Gif-Sur-Yvette Cedex, France\\
        E-mail: \email{matthias.lorentz@cea.fr}}

\author{Pierre Brun\\
         DSM/Irfu, CEA Saclay, F-91191 Gif-Sur-Yvette Cedex, France\\
        E-mail: \email{pierre.brun@cea.fr}}

\author{David Sanchez\\
		LAPP, Université Savoie Mont-Blanc, CNRS/IN2P3, F-74941 Annecy-le-Vieux, France\\
        E-mail: \email{david.sanchez@lapp.in2p3.fr }}

\author{For the H.E.S.S collaboration}

\abstract{When very high-energy photons (VHE, E>100 GeV) propagate over cosmological distances, they interact with background light by pair production. Observations of spectral features in the VHE band of extragalactic sources related to this energy-dependent absorption process with the H.E.S.S. array of Cherenkov telescopes allow measuring the spectral energy distribution (SED) of the extragalactic background light (EBL), otherwise very difficult to determine. Preliminary results on the determination of the SED of the EBL will be presented, based on the measurements of the energy spectra of blazars with H.E.S.S. at redshifts up to $z=0.2$. This model independent approach shows that the shape and overall normalization of the EBL SED is accessible.}

\FullConference{The 34th International Cosmic Ray Conference,\\
		30 July- 6 August, 2015\\
		The Hague, The Netherlands}

\begin{document}

\section{Introduction}

The extragalactic background light ( EBL) is the background radiation field originating from starlight emission and its re-processing by interstellar medium integrated over cosmic history. The spectral energy distribution (SED) of the local EBL (at redshift $z = 0$) contains unique information about galaxy formation and evolution. Direct measurements of this SED are however very difficult because of foreground contamination. Strict lower limits have been derived from galaxy counts, and constraints on the EBL can also be derived using extragalactic very-high energy (VHE) $\gamma$-ray sources. For a review see \cite{Hauser01} and \cite{Dwek13}. Indeed, VHE $\gamma$ rays interact with IR to UV photons via electron-positron pair production, resulting in an attenuated observed flux over the corresponding energy threshold. While this limits the study of extragalactic $\gamma$-ray sources, it also provides a way to probe the EBL over cosmic history. The aim is here to perform a measurement of the EBL density based on $\gamma$-ray spectra.
The optical depth for a VHE photon of energy $E_\gamma$ traveling through a medium with EBL physical density $n( \epsilon, z)$ from a source at $z_s$ is:

\begin{equation}
\tau(E_\gamma,z_s) =c \int_0^{z_s} dz \frac{dt}{dz}  \int_0^2 d\mu \frac{\mu}{2} \int_{\epsilon_{thr}}^\infty d \epsilon \frac{dn(\epsilon,z)}{d\epsilon} \sigma_{\gamma \gamma} \left(E_\gamma(1+z), \epsilon, \mu \right ),
\label{eq:1}
\end{equation}
where $dt/dz=\left( H_0 (1+z) \sqrt{\Omega_M(1+z)^3 +\Omega_\Lambda}\right) ^{-1}$, $\mu=1- \cos(\theta)$ with $\theta$ being the angle between the photon momenta, $\epsilon_{thr} (E_\gamma,z)= \frac{2 m_e^2 c^4}{ E_\gamma \mu(1+z)}$ and $\sigma_{\gamma \gamma}$ is the Bethe-Heitler cross section for pair production~\cite{Gould67}. The absorption effect on the intrinsic spectrum of an extragalactic source is expressed as $\Phi_{obs}(E_\gamma)= \Phi_{int}(E_\gamma) e^{-\tau(E_\gamma,z_s)}$. EBL absorption then leaves a redshift and energy dependent imprint on the observed VHE spectra. To perform a measurement of the EBL SED, a significant detection of this modulation is necessary, which  requires studying spectra with high signal to noise ratio.\\

Different models of the EBL have been developed (see e.g \cite{Fr08}, \cite{Dominguez11}, \cite{Finke10}) and they more or less agree on the overall local EBL shape : an optical to near infrared bump corresponding to starlight emission, a far infrared bump corresponding to the thermal re-emission of starlight by interstellar medium, in particular dust, and a mid infrared "valley" between these two bumps. The exact properties of the energy distribution in the mid infrared is however still not well known and requires more observational input.\\

A previous EBL study with H.E.S.S  \cite{Biteau12} was performed by simultaneously fitting the EBL optical depth and intrinsic spectra of extragalactic sources with a maximum likelihood method. The shape of the EBL SED was fixed {\it a priori} choosing the model of \cite{Fr08}, leaving only the normalization free. The overall test statistics led to a 8.8-$\sigma$ detection of EBL absorption with respect to no absorption, with a normalization factor of $1.27^{+0.18\ }_{-0.15\ } ({\rm stat}) \pm 0.25 ({\rm sys})$.

The new analysis presented here follows a very different approach, focusing on the determination of the shape of the EBL SED in addition to its overall normalization. The sample of blazars used and the data analysis are described in Sec. 2. In Sec. 3, the method to access the shape of the EBL SED is presented. Finally, the results of this preliminary study are shown and discussed in Sec. 4.

\section{H.E.S.S. data}

The high energy stereoscopic system (H.E.S.S.) is an array of five imaging atmospheric Cheren\-kov telescopes located 1800 m above see level in the Khomas Highland, Namibia, detecting $\gamma$ rays ranging from $\sim$ 100  GeV \footnote{For the fist phase of H.E.S.S.. This is brought down to a few tens of GeV with the fifth telescope.}. Due to the peaked pair-creation cross section around the energy threshold, this is precisely the corresponding sensitivity range for EBL absorption from $\sim 0.3 \; \rm \mu m$ to $ \lesssim 100 \; \rm \mu m$. 
In the present study, only data from the first phase of H.E.S.S. with four telescopes are considered.
The sources used are bright H.E.S.S. blazars with a cut on the detection significance  of 10 $\sigma$ : PKS 2005-489 ($z=0.071$),
PKS 2155-304 ($z=0.116$), 1ES 0229+200 ($z=0.14$), H 2356-309 ($z=0.165$), 1ES 1101-232 ($z=0.186$), and 1ES 0347-121 ($z=0.188$). This cut in significance ensures a large number of $\gamma$ rays from the source, a high signal to noise ratio and good reconstruction of energy spectra. The different data sets used are listed in Table \ref{Datasets}. As in \cite{Biteau12}, for variable sources data are divided into smaller subsets restricted in flux range.

 The main analysis is performed using Model analysis~\cite{2009APh....32..231D}, and the spectra were reconstructed using an unfolding technique~\cite{2007NIMPA.583..494A}. The covariance matrix determined during the unfolding procedure is used in order to take into account the  correlations between bins. A cross-check analysis is performed with ImPACT analysis~\cite{2014APh....56...26P} and an independent calibration chain, yielding consistent results.\\

\begin {table}[H]
\begin{center}
\begin{tabular}{|c|c|c|}
  \hline
 Data set & $z$ & $E_{min} - E_{max} \ (TeV)$ \\
  \hline
  1ES 0347-121 & $0.188$ & $0.20 - 6.5$ \\
  1ES 1101-121 & $0.186$ & $0.21 - 8.8$ \\
  H 2356-309 & $0.165$ & $0.19 - 10.3$\\
  1ES 0229+200 & $0.14$ & $0.69 - 10.2$\\
  PKS 2155-304 (1) & $0.116$ & $0.23 - 10.1$\\
  PKS 2155-304 (2) & $0.116$ & $0.21 - 9.6$\\
  PKS 2155-304 (3) & $0.116$ & $0.21 - 12.0$\\
  PKS 2155-304 (4) & $0.116$ & $0.21 - 11.1$\\
  PKS 2155-304 (5) & $0.116$ & $0.21 - 6.0$\\
  PKS 2155-304 (6) & $0.116$ & $0.26 - 15.0$\\
  PKS 2155-304 (7) & $0.116$ & $0.26 - 13.9$\\
  PKS 2155-304 (8) & $0.116$ & $0.21 - 12.9$\\
  PKS 2005-489 (1) & $0.071$ & $0.26 - 8.9$\\
  PKS 2005-489 (2) & $0.071$ & $0.26 - 11.1$\\
   \hline
\end{tabular}
\end{center}
\caption{Data sets used in this study with redshift and energy range covered by the observed $\gamma$-ray spectra.}
\label{Datasets}
\end{table}

\section{Method}

\subsection{Accessing the EBL shape}

The local, $z=0$ EBL SED is described by splines constructed from a logarithmic grid in the ($\lambda, \ \nu F_\nu$) plane in order to test a large variety of EBL shapes. 
Each spline is defined by knot points on the grid as in \cite{Mazin07, Meyer12}. 
All tested shapes comply with the strict lower limits from galaxy counts (see references in \cite{Biteau15, Dwek13}), while the maximum is set relatively high over current upper limits in order to be conservative. Two shifted grids against each other are used to reduce the constraints on shapes imposed by the positions of the knots. Each shape is used to fit the data and is associated to a goodness-of-fit estimator in the form of a $\chi^2$. 
The range covered by the grids covers the range of H.E.S.S. $\gamma$-ray absorption sensitivity ($\sim$ 100 GeV - 20 TeV). In total 116,640 EBL shapes were tested.

\begin{figure}[H]
\center
\includegraphics[scale=0.35]{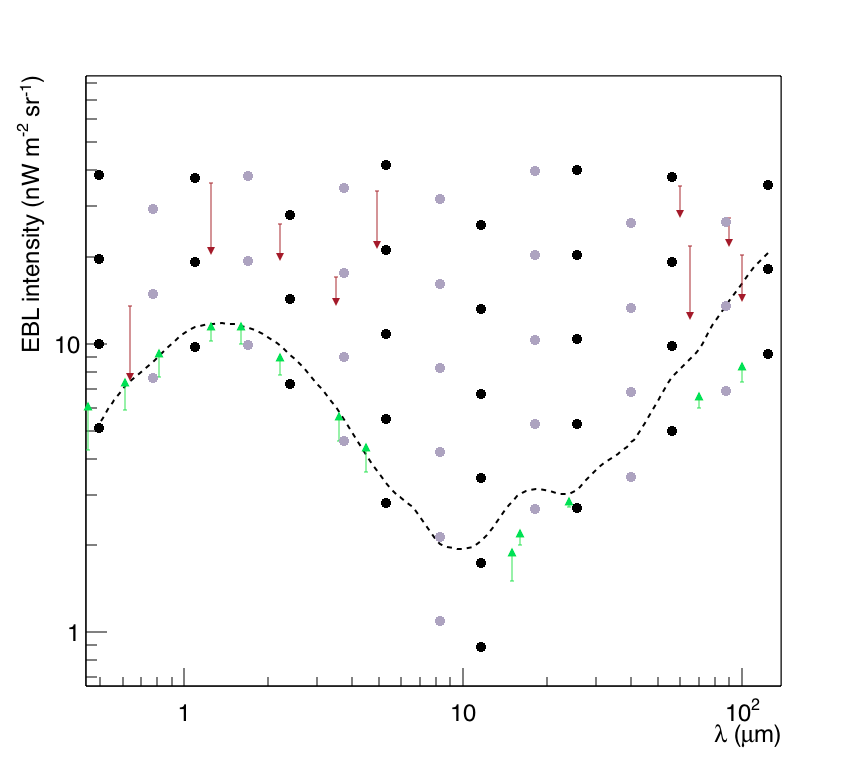}
\caption{\label{EBLGrids} Grid setup to construct EBL shapes. The black and grey dots are the knots for the two shifted grids. Also shown are lower limits from galaxy counts in green,  upper limits from direct measurements in red extracted from \cite{Dwek13}, \cite{Biteau15}. The dotted curve is the reference local EBL from the model of \cite{Fr08}.}
\end{figure}

For each spectrum, a fit with EBL absorption is performed considering every shape $i$ of the two grids, $\Phi_{obs}(E_\gamma)= \Phi_{int}(E_\gamma) e^{-\tau_i(E_\gamma,z_s)}$. The intrinsic (de-absorbed) VHE spectra are fitted with a power law and a log parabola, the best fit of which is selected considering the lower $\chi^2$.  
Further improvements of the analysis will include consideration of other spectral shapes. 
%
%
%
%
The optical depth depends not only on the local $z=0$ EBL density, but also on its evolution with redshift. This intrinsic evolution of the EBL is usually taken into account by introducing an additional scaling index to the cosmological photon density in the form of $(1+z)^{f_{evol}}$, with $f_{evol}$ ranging from 1.2 to 1.7~\cite{Meyer12, Biteau15}.
Another approach is used here by taking the ratio of the density at redshift $z$ to the density at redshift $z=0$ from the model of~\cite{Fr08}. It was checked that this does not introduce a strong model-dependence since a comparison with a fixed scaling index did not yield significant differences. Further improvements of the analysis will include different hypotheses.


\subsection{Combination of several spectra}

At first approximation, the EBL photon field should be essentially identical to all sources \cite{Furniss15}.
Since each source provides a different combination of redshift and energy range, the results from the spectra of different sources can be combined to determine the EBL in a more significant way than using a single spectrum and over a wider range of energy, as a collective signature of EBL absorption.

For each data set the envelope of all $z=0$ EBL shapes inside the $\chi^2_{min} +1$ interval is retrieved. This envelope is considered as a $1-\sigma$ statistical contour of an actual measurement of the EBL SED. 
Note that this measurement is made at specific wavelengths defined by the grid spacing. The combination of the measurements from each data set is done as a combination of $\chi^2$ from independent measurements. This way, the combined central value for each wavelength is the result of the average of the measurements from each data set weighted by the corresponding $1-\sigma$ values.


\section{Results and discussion}

\begin{figure}[H]
\center
\includegraphics[scale=0.41]{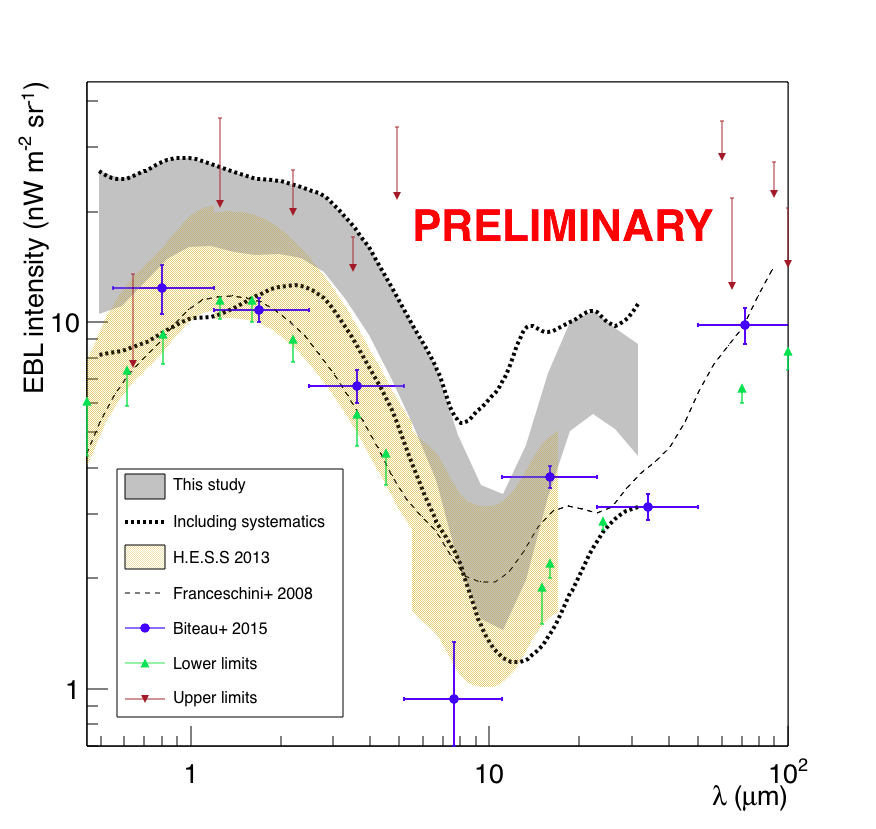}
\caption{\label{FinalBand} Preliminary result of this work together with previous results, limits and the model of \cite{Fr08}}
\end{figure}

The result of this preliminary work to access the EBL shape is summarized in Fig.~\ref{FinalBand}. In that figure, the grey area shows the combined best-fit EBL SED shape from all considered data sets, compared to previous measurements and constraints, as well as the fiducial model of~\cite{Fr08}. In order to be conservative the wavelength range presented here have been restricted to match the pair creation threshold with the lowest of the last points in energy from our data sample, i.e 6.5 TeV. Considering the highest points of the data sample would lead to a further coverage in the infrared.

When comparing with previous measurements of the EBL density by H.E.S.S. it is important to note that the assumptions made are very different. Indeed in~\cite{Biteau12} the $z=0$-shape and evolution of the fiducial model \cite{Fr08} was assumed and only the normalization was set free and fitted to the data. Here no prior is used on the considered $z=0$ EBL shape and still the shape determined with our method is relatively close to the model of~\cite{Fr08}. In Fig.~\ref{FinalBand}, the dotted grey lines indicate the level of systematic uncertainties on the shape determination. Those have been obtained using different calibrations, analysis chains and spectral reconstructions. Part of the width of the systematic uncertainty comes from a jackknife analysis where some data sets were removed one by one. The final grey shaded area in Fig.~\ref{FinalBand} is drawn including all data sets. The main analysis leads to an EBL intensity around 20 $ \rm nW  \ m^{-2} \ sr^{-1}$ (at the optical peak) slightly higher than previous measurements, at the level of upper limits. However, this excess is not significant given our current understanding of systematics.

In the near future, improvements of the analysis will include consideration of other intrinsic spectral shapes. It is foreseen to include more sources at various redshifts and consider different hypotheses regarding the evolution of the EBL SED. Using spectra obtained with the second phase of H.E.S.S. with a lower energy threshold will also provide more leverage to constrain the short wavelength range of the EBL.

\section{Conclusions}

The results presented here are the first measurement of the local extragalactic background light SED {\it shape} with H.E.S.S., where previous studies dealt with its overall normalization. These preliminary results are based on first-phase H.E.S.S. data, and show that  the sensitivity of the method allows to find results comparable to fiducial models in terms of shape, and broadens the previous H.E.S.S. measurements in terms of EBL wavelengths.

This analysis is a first step towards a precise determination of the EBL SED with $\gamma$-ray absorption using H.E.S.S. data. This is a prerequisite for analyses aiming at the determination of the intrinsic VHE spectra of blazars and it is also closely related to alternative studies where the effective mean free path of TeV $\gamma$ rays is modified (new fundamental physics like axions,  secondary emission from cascades in the intergalactic medium, gravitational lensing, ...).\\

\section*{Acknowledgement}
The support of the Namibian authorities and of the University of Namibia in facilitating the construction and operation of H.E.S.S. is gratefully acknowledged, as is the support by the German Ministry for Education and Research (BMBF), the Max Planck Society, the German Research Foundation (DFG), the French Ministry for Research, the CNRS-IN2P3 and the Astroparticle Interdisciplinary Programme of the CNRS, the U.K. Science and Technology Facilities Council (STFC), the IPNP of the Charles University, the Czech Science Foundation, the Polish Ministry of Science and Higher Education, the South African Department of Science and Technology and National Research Foundation, and by the University of Namibia. We appreciate the excellent work of the technical support staff in Berlin, Durham, Hamburg, Heidelberg, Palaiseau, Paris, Saclay, and in Namibia in the construction and operation of the equipment.\\
Part of this work is supported by the French Agence Nationale de la Recherche as part of the CosmoTeV project, and by the French high-energy astrophysics national program PNHE. D.S. is supported by the Labex ENIGMASS.

\end{document}